\documentclass[%
 aip,
 jmp,%
 amsmath,amssymb,
 preprint,%
]{revtex4-1}

\usepackage[latin1]{inputenc}
\usepackage{graphicx}
\usepackage{amsmath}
\usepackage{amssymb}
\usepackage{color}
\usepackage{bm}
\usepackage{epsfig}
\usepackage{tikz}
\usepackage{setspace}
\DeclareRobustCommand*\circled[1]{\tikz[baseline=(char.base)]{
            \node[shape=circle,draw,inner sep=2pt] (char) {#1};}}

\newcommand*{\rom}[1]{\expandafter\@slowromancap\romnumeral #1@}

\begin{document}

\preprint{APS/123-QED}

\title{Vortices catapult droplets in atomization}

\author{J John Soundar Jerome}
	\email{soundar@dalembert.upmc.fr}
\affiliation{%
 UPMC Univ. Paris 06 \& CNRS - UMR 7190, Institut Jean Le Rond d'Alembert, F-75005 Paris, France
}%
\author{Sylvain Marty}
\author{Jean-Philippe Matas}
\affiliation{%
Univ. Grenoble Alpes, Laboratoire des \'{E}coulements G\'{e}ophysiques et Industriels (LEGI), F-38000 Grenoble, France
}%

\author{St\'{e}phane Zaleski}
\author{J\'{e}r\^{o}me Hoepffner}
\affiliation{%
 UPMC Univ. Paris 06 \& CNRS - UMR 7190, Institut Jean Le Rond d'Alembert, F-75005 Paris, France
}%

\date{}

\begin{abstract}
A droplet ejection mechanism in planar two-phase mixing layers is examined. Any disturbance on the gas-liquid interface grows into a Kelvin-Helmholtz wave and the wave crest forms a thin liquid film that flaps as the wave grows downstream. Increasing the gas speed, it is observed that the film breaks-up into droplets which are eventually thrown into the gas stream at large angles. In a flow where most of the momentum is in the horizontal direction, it is surprising to observe these large ejection angles. Our experiments and simulations show that a recirculation region grows downstream of the wave and leads to vortex shedding similar to the wake of a backward-facing step. The ejection mechanism results from the interaction between the liquid film and the vortex shedding sequence: a recirculation zone appears in the wake of the wave and a liquid film emerges from the wave crest; the recirculation region detaches into a vortex and the gas flow over the wave momentarily reattaches due to the departure of the vortex; this reattached flow pushes the liquid film down; by now, a new recirculation vortex is being created in the wake of the wave---just where the liquid film is now located; the liquid film is blown-up from below by the newly formed recirculation vortex in a manner similar to a bag-breakup event; the resulting droplets are catapulted by the recirculation vortex.
\end{abstract}


\maketitle

\section {Introduction}
\label{sec:Intro}
	Atomisation is the process by which a liquid stream fragments into droplets. It is a common phenomenon in nature and industry (for instance, see Chapter $1$ in \textit{Atomization and Sprays} by \citet{Lefebvre_1989}). One of the ways to make droplets or sprays is to form waves on the gas-liquid interface by a fast-moving gas on a liquid surface, for example, air-blast injectors systems. The textbook \textit{Liquid Atomization} by \citet{Bayvel_n_Orzechowski_1993} provides a detailed study of such injector systems. The waves on the gas-liquid interface grow by extracting the kinetic energy of the liquid and gas stream and if the kinetic energy is sufficiently large, thin liquid sheets or films are formed which break into droplets \citep{Eggers_1997}. This step is called primary atomization. During the final and secondary atomization, these droplets form a fine spray via collision and stretching. While the latter process determines the size and distribution of the droplets, the former plays an important role in determining the rate at which droplets are produced and the initial conditions for the extent of the dispersed two-phase flow. The physical mechanisms of primary atomization are often complex, nonlinear and hence, are poorly understood. This is true not only for co-flowing gas-liquid mixing layers but also for jets \citep{Lin_n_Reitz_1997, Lasheras_n_Hopfinger_2000, Eggers_n_Villermaux_2008}, planar sheets \citep{Duke_2012}, etc. In this article, the primary atomization process in a co-flowing gas-liquid mixing layer is illustrated, in particular, when the horizontal gas flow is fast.
	
	\citet{Rayleigh_1879} showed that in a single-phase mixing layer, the Kelvin-Helmholtz ($KH$) type instability wavelength and growth rate is directly related to the thickness of the shear layer. In the case of two phase mixing layers, thanks to a large body of experimental evidence \citep{Villermaux_thesis_1993, Raynal_1997, Villermaux_1998, Raynal_thesis_1997, Rayana_2006, Rayana_thesis_2007, Matas_2011}, it is now well-established that the instability wavelength is governed by the gas boundary layer thickness $\delta_{g}$. Combining both experimental and numerical investigations, \citet{Otto_2013} \& \citet{Fuster_2013} show that depending on the momentum ratio $M = {\rho_{g}U_{g}^{2}}/{\rho_{l}U_{l}^{2}}$ (where $\rho_{g}$, $\rho_{l}$ represent the gas and liquid density and $U_{g}$,  $U_{l}$ represent the gas and liquid freestream velocity), such an instability leads to a noise amplifier or a nonlinear global mode \citep{Huerre_n_Rossi_1998} that beats at a particular frequency. A two-stage mechanism for interface destabilization has been demonstrated by \citet{Marmottant_2004} for co-axial gas-liquid jets. They showed that, at first, the  instability leads to waves whose length scale is directly related to the gas boundary layer thickness and the density ratio.  Later, the transient acceleration of the liquid surface induced by the waves can promote a Rayleigh-Taylor type instability at the wave crests forming liquid ligaments. A similar two-step mechanism is also put forward for the case of planar two-phase mixing layers by \citet{Hong_2003} who proposed that the transient accelerations due to the primary destabilization (Kelvin-Helmholtz instability) should be modified to account for the aerodynamic acceleration of thin ligaments due to the drag exerted by the air flow in the horizontal direction. On the other hand, optimal growth studies in two-phase mixing layers \citep{Yecko_n_Zaleski_2005} also suggest that ligament formation could be related to  large transient growth resulting in strong liquid up-flows and high-speed streamwise gas jets near the interface. However, \citet{Boeck_2007} later showed via direct numerical simulations that relatively large Reynolds and Weber numbers  are necessary to observe the nonlinear development of perturbations into growing ligaments. Despite the evidence for $3D$ dimensional structures in planar mixing layer experiments \citep{Hong_2003}, there is relatively good agreement between linear stability analysis based on parallel flow assumptions. For instance, by taking into account the liquid velocity deficit at the gas-liquid interface, \citet{Matas_2011} demonstrated a good agreement of the measured frequency with the frequency predicted by the inviscid stability analysis. Similarly, \citet{Otto_2013} and \citet{Fuster_2013} also provided relatively good comparisons between experiments and viscous linear spatio-temporal stability results.
	
	In the present work, the various mechanisms of such interfacial pattern formations are not considered. However, the interaction between these interfacial patterns and the gas flow field is particularly analysed. For example, bag break-up is known to occur in round liquid jets exposed to a gas flow at gas Weber numbers (based on the diameter of the jet and the gas speed) less than $30$. The jet first deforms into a curved sheet due to aerodynamic drag, followed by the formation of one or more bags, along the jet-streamwise direction. These bags expand and ultimately burst. A detailed account on the formation and breakup of such bags is given by \citet{Sankarakrishnan_Sallam_2008} Recently, \citet{Scharfman_n_Techet_2012} identified multiple bag-breakup in such flows when the jet diameter is larger than the capillary length of the liquid. One can also expect a strong interaction between gas-liquid interfacial patterns and the gas flow. Such interactions determine how droplets are created in the primary atomization and so, the initial droplet distribution to determine the final dispersed state via the secondary atomization. Hence, it is central to understand such processes.

\begin{figure}
\begin{center}

\epsfig{file=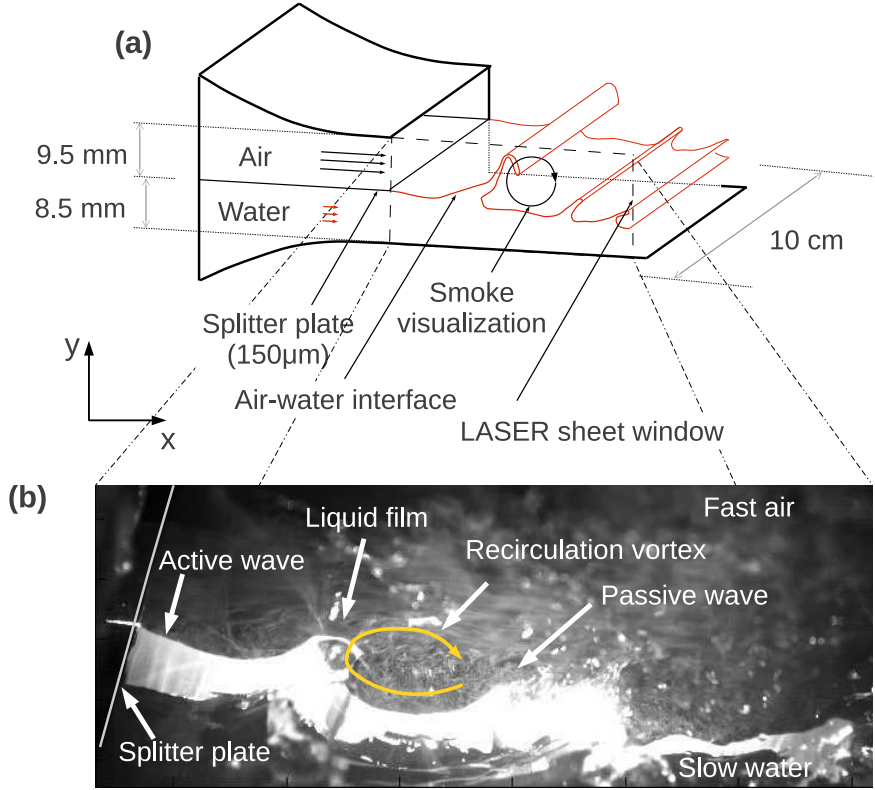,width=0.7\textwidth,keepaspectratio=true,bbllx=145,bblly=383,bburx=459,bbury=720,clip=}

\end{center}
\caption{(a) Schematic of the planar air-water mixing layer from the experimental set-up at LEGI. (b) Airborne smoke particles and a strong LASER sheet are used to observe $2D$ structures, namely, the liquid film and the recirculation zone downstream of the wave (see \textit{Movie 2} of the supplementary material \cite{SuppMovie} for more details).}.
\label{fig:Schematic}
\end{figure}

\begin{figure}
\begin{center}

\epsfig{file=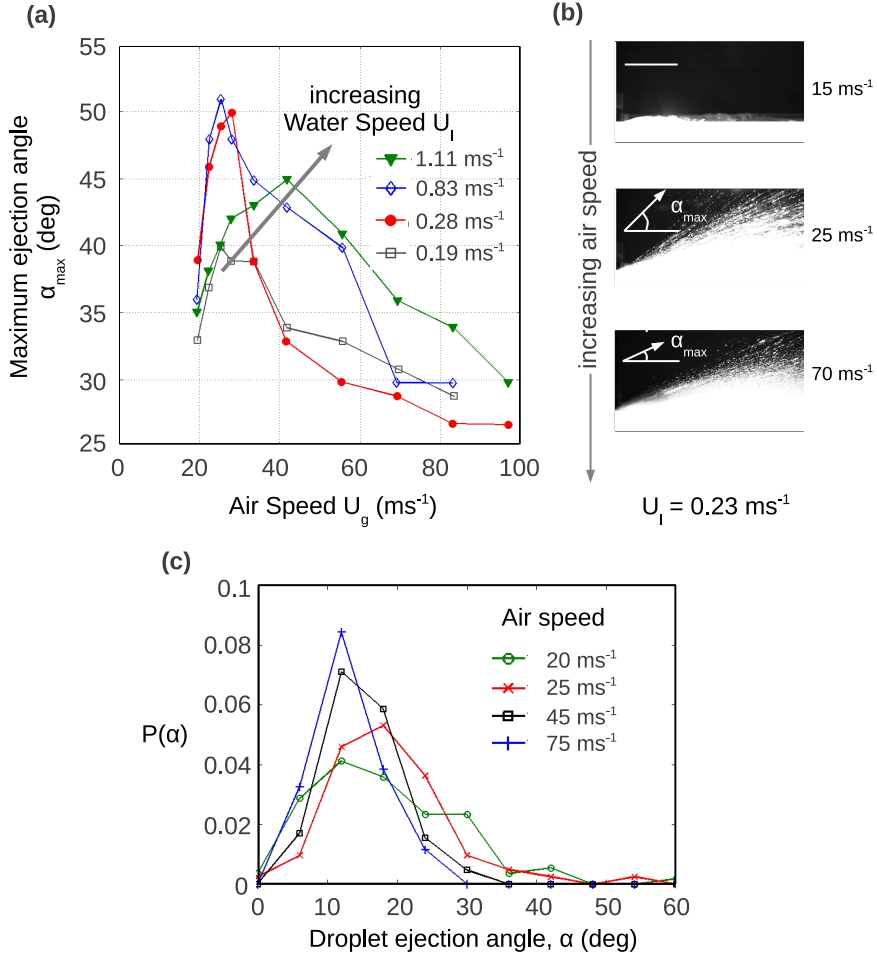,width=0.7\textwidth,keepaspectratio=true,bbllx=144,bblly=395,bburx=429,bbury=708,clip=}

\end{center}
\caption{(a) Maximum droplet ejection angle $\alpha_{max}$ as a function of air speed $U_{g}$. The data is extracted from figures $3.25$ (page $113$) of \citet{Raynal_thesis_1997}. (b) Recent mixing layer experiments at LEGI for a water speed of $0.23 ms^{-1}$: superposition of air-water interface at successive times are shown. We observe that droplets are ejected starting from a critical air speed and $\alpha_{max}$ varies non-monotonically. (c) The probability distribution function $P(\alpha)$ for various ejection angles and airspeeds at a fixed water speed of $0.23 ms^{-1}$. We find that large ejection angles are not rare events.}
\label{fig:AngleExperimental}
\end{figure}

	Consider for example figure \ref{fig:Schematic}b which shows interfacial patterns and complex gas flow structures during the atomization process in a planar two-phase mixing layer. It is taken by a high-speed camera (Photron SA1.1) in the splitter plate experimental set-up illustrated in the schematic (figure \ref{fig:Schematic}a). The set-up uses a Argon LASER sheet across an air-water mixing layer past a thin splitter plate in order to observe the two-dimensional structures in the flow (see \citet{Matas_2011} for more details on the experimental set-up). The gas flow is visualized using airborne smoke particles while fluorescein is mixed with water so that the air-water interface can be  distinctly captured under LASER sheet illumination. In figure \ref{fig:Schematic}b, the LASER sheet illuminates the liquid surface showing two waves: the active wave which grows while remaining attached to the splitter plate and the passive wave which is the previous active wave that has left the plate. Between these waves, there is a recirculation region and just above it, a liquid film is readily visible. As the liquid film develops, it flaps and droplets are violently extracted from the crest of the active wave. \citet{Raynal_thesis_1997} carried out measurements of the maximum droplet ejection angles $\alpha_{max}$ on an older version of the same experimental set-up. Figure \ref{fig:AngleExperimental}a displays the variation of $\alpha_{max}$ as a function of air speed at four different liquid velocities. When the air speed $U_{g}$ is increased progressively, the angle $\alpha_{max}$ increases steeply until about a critical value (as large as $50^\circ$) and then decreases monotonically, however slowly, with further increase in $U_{g}$ (see also \citet{Rayana_thesis_2007}). When the liquid speed is increased the same behaviour is observed, but the maximum angles are shifted to larger values. \citet{Raynal_1997} used superposition of images (for example, see figure \ref{fig:AngleExperimental}b) in order to measure the maximum ejection angles given in figure \ref{fig:AngleExperimental}a. We see clearly that the ejection angles vary non-monotonically with air speed.
	
	In order to gain insight into the statistics of ejection angles, we measured the angle of ejection $\alpha$ by carefully observing individual ejection events. We carried out measurements of individual ejection angles for a fixed liquid velocity $U_{l} = 0.23 ms^{-1}$ at four different air speeds. For each of these conditions, about $50$ ejection events are identified and analysed to build an approximate probability density function $P\left(\alpha\right)$. It is presented in figure \ref{fig:AngleExperimental}c. We observe that maximum values of $\alpha$ reach up to $50$ degrees for $U_{g} = 20  ms^{-1}$ and $U_{g} = 25  ms^{-1}$, but decrease down to $25$ degrees for $U_{g} = 70  ms^{-1}$. This trend is consistent with the data of figure \ref{fig:AngleExperimental}a since these maximum angles should fall between the data of $U_{l} = 0.19  ms^{-1}$ and $U_{l} = 0.28  ms^{-1}$. Note that figure \ref{fig:AngleExperimental}c clearly indicates that large ejection angles are not rare events: for $U_{g} = 25  ms^{-1}$, around $40$\% of ejection events correspond to maximum ejection angles larger than $20^{\circ}$.
	
	Thus, we observe that droplets are thrown into the air stream at a considerable angle with respect to the horizontal axis. In a flow system with large horizontal momentum, it is intriguing to find that droplets move in an oblique trajectory. The aim of the present work is to understand the physical mechanisms causing such a phenomenon.
	
	The experimental results of \citet{Raynal_thesis_1997} in figure \ref{fig:AngleExperimental}a correspond to the case where the velocity of the air-flow (the lighter fluid) is large compared to that of water (the heavier fluid) with an air-flow recirculation region as identified in figure \ref{fig:Schematic}b. The interaction of this zone with the  wave crest and hence its influence on primary atomization processes have rarely been considered. However, similar situations in which air-flow separation and the resulting recirculation region play a significant role are well-known for the case of wind-induced waves in ocean: wave breaking and ``freak'' waves (``rogue'' waves or extreme wave events). \citet{Jeffreys_1925, Jeffreys_1926} showed that surface waves in the ocean are formed mainly due to the pressure difference created by the air-flow over the water surface. Wave breaking corresponds to the initial stage of overturning motion of the wave crest that creates sea-sprays (even jets in most cases) and foams \citep{Peregrine_1983}. The presence of air-flow separation during wave breaking was shown by \citet{Banner_n_Melville_1976} and \citet{Banner_1990}. Later, \citet{Reul_1999} described the instantaneous velocity field of separated air-flow over breaking waves. It is now recognised that air-flow separation over breaking waves enhances momentum transport from air to water \citep{Melville_2002, Alexakis_2004, Reul_2008, Sullivan_n_McWilliams_2010}. 
	
	During the last decade, considerable work had been done to throw light upon the importance of such flow separation on freak waves (giant waves appearing sporadically on the sea surface). \citet{Touboul_2006} and \citet{Kharif_2008} showed that the time duration of freak wave mechanism is increased by the presence of a recirculation region behind the wave. They also demonstrated an increase in the freak wave height. It is thus expected that the recirculation vortex can show strong interactions with the  wave crest and hence, play an important role in the atomization process. However, the influence of vortices on the dynamics of gas-liquid interface is not well known, largely due to the fact that such events are complex and involve a large variety of scales. It is precisely the objective of this work to demonstrate such vortex-interfacial wave interactions.

\section {Vortex shedding as a driving mechanism for droplet ejection}
\label{sec:VortexSheddingExpeObserv}

\begin{figure}
\begin{flushleft}

\epsfig{file=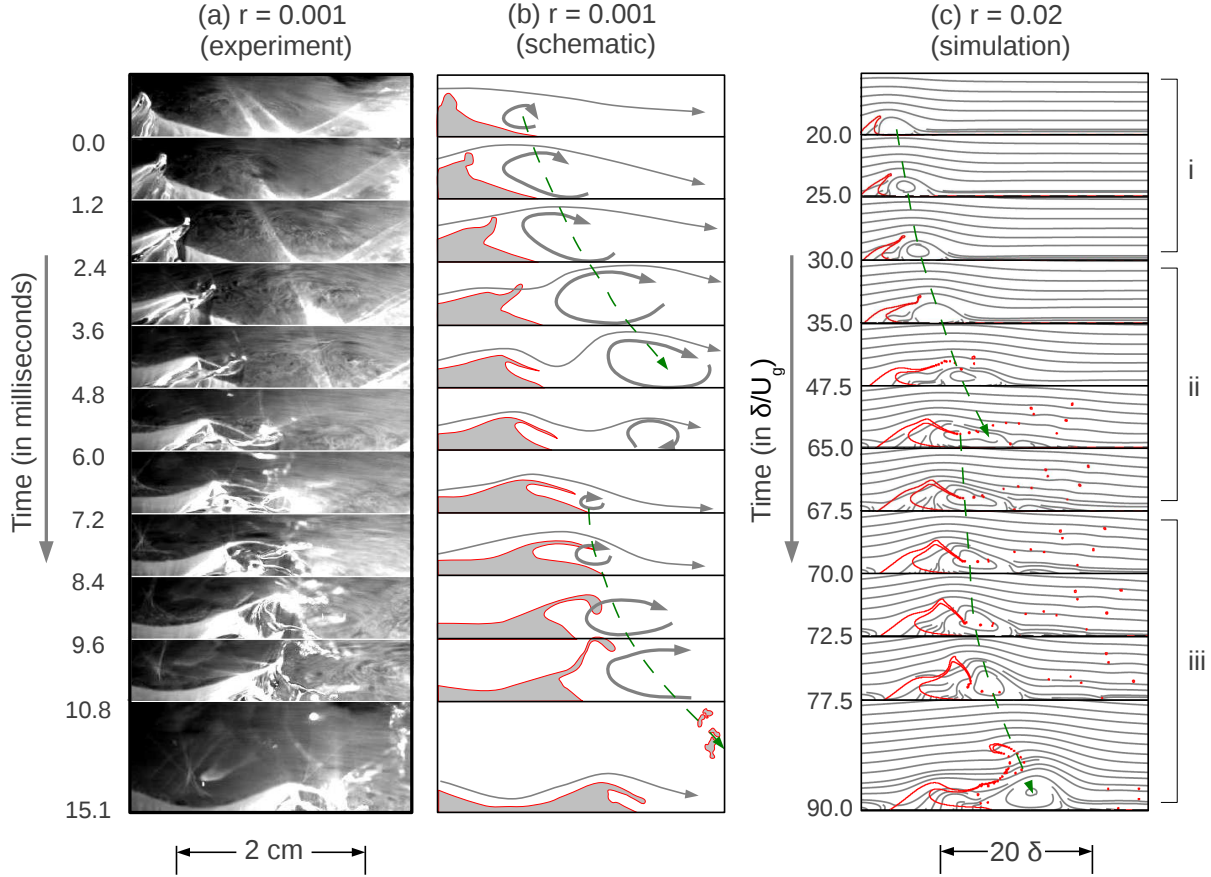,width=0.55\textwidth, keepaspectratio=true}

\end{flushleft}
\caption{The sequence of the vortex catapult mechanism, shown experimentally, schematically and numerically. (a) Snapshots of air-water interface and air flow visualisation at $U_{g} = 25.5$ $ms^{-1}$ and $U_{l} = 0.23$ $ms^{-1}$, $r=0.001$. (b) Schematic of the air-water snapshots. (c) Snaptshots of the gas-liquid interface and streamlines from Direct Numerical Simulation of a single nonlinear Kelvin-Helmholtz wave at $r=0.02$. The numbering pertains to the catapult sequence as described in the text. See our supplementary material \cite{SuppMovie} for \textit{Movies 1 $\&$ 2}.}
\label{fig:ExpeVortexShedding}
\end{figure}

A series of snapshots of the air-water interface and smoke visualisation of the airflow past it is displayed in figure \ref{fig:ExpeVortexShedding}a. In order to render the interface and the flow visualization more visible, a scale-to-scale schematic of these snapshots is given in figure \ref{fig:ExpeVortexShedding}b. Here, thick lines with arrows represent the air flow and, in particular, the emphasis is put on the recirculation vortex (see \textit{Movies 1} \& \textit{2} in the supplementary video \cite{SuppMovie} for more details). The interface (red online) and air flow evolve as we march down the time axis from top to bottom in both figures. As the wave grows in amplitude, a thin liquid film is formed. Smoke visualization shows the presence of a separated flow with a recirculation zone just below the liquid film. The recirculation zone grows in time and blows upward on the liquid film above it. Note that the size of the zone compares with the height of the wave. The recirculation vortex is eventually shed. During the entire process, the wave moves much slower than the air stream and hence, acts as an obstacle to the air flow. This implies that the air flow past the wave is similar to the flow past a backward facing step (analogous to the case of breaking waves \citep{Reul_1999} and freak waves \citep{Touboul_2006}). This is the reason why the air flow over the interface wave separates and the separated flow reattaches after a small recirculation zone in figure \ref{fig:ExpeVortexShedding}. 

We may now proceed to the detailed description of the events shown in the figure. We may call this sequence the \textit{droplet catapult mechanism}:
\begin{center}
\begin{enumerate}
\renewcommand{\theenumi}{\roman{enumi}}
\item A recirculation appears in the wake of the wave and a liquid film emerges from the wave crest. 
\item The recirculation region detaches into a vortex. The departure of the recirculation vortex leads to a momentary reattachment of the gas flow along the wave. This reattached flow, in turn, pushes the thin liquid film downward. 
\item A new recirculation region appears in the wake of the wave---precisely where the liquid film is now located. Thus, the liquid film is, eventually, blown-up from below by the nascent recirculation vortex, similar to a bag-breakup event. The resulting droplets are catapulted by the shed vortex.
\end{enumerate}
\end{center}

We refer the reader to \textit{Movie 2} of the supplementary material \cite{SuppMovie} where these droplet catapult events via vortex shedding are shown for various air speeds. We observe that the droplets that are ejected at large angles originate from the liquid film growing at the crest of the wave. It is clear from \textit{Movie 2} \cite{SuppMovie} that the thin liquid film is blown-up from \textit{below} by none but the recirculating air flow. We, hereafter, refer to this break-up as \emph{bag-breakup from below} whereby a thin liquid sheet attached to a liquid rim breaks-up (similar to a soap film attached to a ring) as it is blown-up by the recirculation vortex into a bag. Bag-breakup is well-known to be a violent event, see for instance \citet{Pilch_n_Erdman_1987, Villermaux_2007, Sankarakrishnan_Sallam_2008, Scharfman_n_Techet_2012}. It also leads to a wide distribution of droplet sizes. This is shown for instance in \citet{Villermaux_2009} where it is the key element to explain the statistics of raindrops. 


\section {Localized self-similar  wave}
\label{sec:SelfSimKHwave}

It is clear from figures \ref{fig:ExpeVortexShedding}a$-$b and supplementary videos (\textit{Movies 1}--\textit{4} \cite{SuppMovie}), that the air-flow visualization of such a shedding process is difficult because of the $3D$ nature of the two-phase mixing layer due to the influence of capillary waves and side walls. These effects mask the visualization of the vortex behind the wave. Moreover, due to the presence of a large number of droplets during the droplet catapult process, it is cumbersome to identify the air-water interface using the LASER sheet as it is reflected unequally by the droplets. Direct Numerical Simulations ($DNS$) can be used to visualise the gas and liquid flow fields. However, $DNS$ computations of such complex three dimensional two-phase flows at the experimental density ratios, gas and liquid speeds are currently not feasible. Thus, we consider, instead, the evolution of a localized initial disturbance in an infinite two-phase $2D$ mixing layer. This study, as we shall see later, illustrates that the \textit{droplet catapult mechanism} can be identified in other two-phase flow configurations as well. Direct numerical simulations of this two--fluid system is implemented via the open source Gerris Flow Solver. A finite volume scheme is used to discretize the incompressible Navier--Stokes equations whereas the interface is traced in the framework of the Volume Of Fluid ($VOF$) method via a quadtree adaptive grid refinement. \citet{Popinet_2003} and \citet{Popinet_2009} provide a comprehensive description of this numerical technique. For a detailed review on various numerical methods in free-surface and interfacial flows, we refer the reader to \citet{Scardovelli_Zaleski_1999, Tryggvason_2011}
	
	An initial impulse disturbance in such flows eventually develops into a nonlinear Kelvin-Helmholtz wave that grows and propagates downstream in a self-similar manner, see for example \citet{Hoepffner_2011} and \citet{Orazzo_2012}. This flow situation is a simple configuration whereby the catapult mechanism in a planar two-phase mixing layer can be readily examined numerically. Here, only the dynamics of the active wave and the effect of fast gas flow are investigated while the role of the passive wave, the splitter plate dimensions, the boundary layer thickness of the incoming flow and gravity are neglected. 
	
	Our numerical investigation consists of an infinite two-phase $2D$ mixing layer with a fast gas flow (density $\rho_{g}$) on top of a liquid at rest (density $\rho_{l}$). Sufficiently far away from the gas-liquid interface, the gas flows at a speed $U_{g} = 1$ in the horizontal x-direction. The viscosity of the two fluids is taken to be the same. Thus, the initial velocity field in the liquid and gas streams is built from error functions that satisfy the stress continuity at the interface. The non-dimensional parameters that characterize this analysis are, namely, the Reynolds number $Re = U\delta/\nu$ where $\delta$ and $\nu$ are the mixing layer thickness and the kinematic viscosity, respectively, and the Weber number $We = \rho_{g}U_{g}^{2}\delta/\sigma$ where $\sigma$ is the surface tension of the liquid. For our simulations, we chose large enough  Reynolds and Weber numbers ($Re = 100$ and $We = 1000$) so that they do not play a deciding role on the droplet catapult phenomenon.
	
	The size of the numerical domain is $500\delta$ in length (x-direction) and $250\delta$ in height (y-direction). Simulations are performed with periodic boundary conditions in the streamwise direction and symmetry boundary conditions at the top and bottom boundaries. The initial condition for the computations consists of a small amplitude vertical impulse disturbance of extent $\delta$ in the x-direction. The amplitude of this disturbance is kept small enough so that it does not create a vertical jet but initiates, instead an isolated nonlinear Kelvin-Helmholtz wave. This initial amplitude is large enough so that it can bypass linear growth of disturbances into a packet of waves. Several spatial discretization levels were tested to validate the results and a mesh size of approximately $0.06 \delta$ units is chosen for which the error in the location of the wave is found to be lesser than $1\%$.
	
	If one neglects viscosity and capillarity (large Reynolds and Weber numbers), the only length scales are $U_{g}t$ and $\delta$. After a short transient, the initial impulse disturbance grows larger than the thickness of the mixing layer $\delta$.  If the vorticity field $\omega$ is considered as a function of $x$, $y$, $t$, $U_{g}$ and $\delta$, at sufficiently large time $t >> \delta/U_{g}$,  it can be shown that  $\omega = U/\delta f\left(x/U_{g}t, y/U_{g}t, \rho_{g}/\rho_{l}\right)$ (see \citet{Hoepffner_2011}). Hence, in the self-similar coordinates $x^{'} = x/U_{g}t$ and $y^{'} = y/U_{g}t$, the shape, size and the dynamics of the wave depends only on a single parameter, namely, the density ratio $r =  \rho_{g}/\rho_{l}$.

\begin{figure}
\begin{center}

\epsfig{file=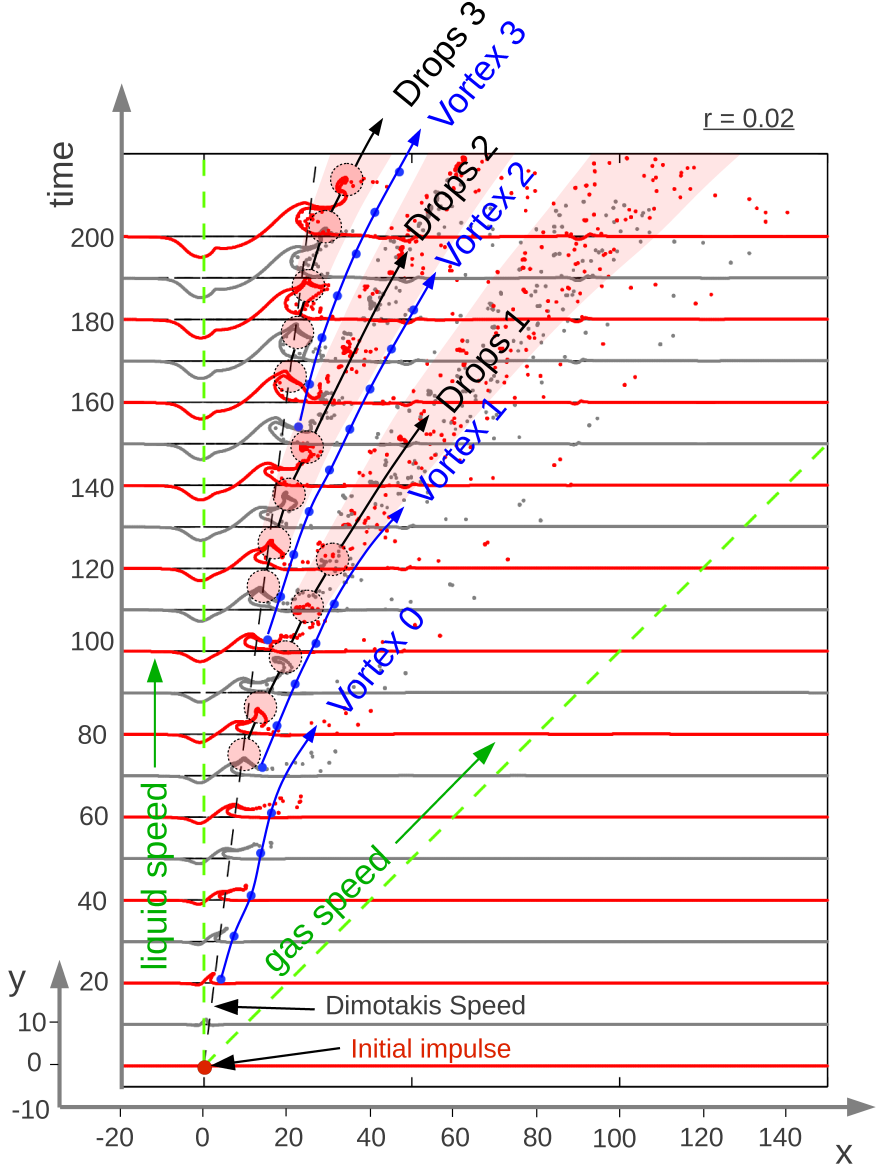,width=0.7\textwidth,keepaspectratio=true}

\end{center}
\caption{Spatio-temporal evolution of the gas-liquid interface for density ratio $r = 0.02$ as obtained from Direct Numerical Simulations. Here, we display three successive droplet catapult sequences. During each sequence, the liquid film at the crest of the wave flaps, bulges-out and breaks-up.}
\label{fig:BagBreakUpSimu}
\end{figure}
	
	The evolution of such a nonlinear self-similar Kelvin-Helmholtz wave due to a localized initial disturbance is shown in figure \ref{fig:BagBreakUpSimu}. It corresponds to the case $r = 0.02$. The time axis is specified in ${\delta}/{U_{g}}$ units. The thick lines (mid-gray and gray lines; see color online) denote the gas-liquid interface wave whose crest forms a liquid film which stretches, bulges-out and eventually breaks-up into droplets due to the presence of shear. Such a wave grows linearly in time as illustrated by \citet{Hoepffner_2011} and it periodically creates droplets. As already observed in figures \ref{fig:ExpeVortexShedding}a$-$b, the wave moves much slower than the gas particles. Note that, unless otherwise mentioned, gas particles do not refer to any tracers but \textit{material} elements of fluid \cite{Batchelor_1967}. Its speed is approximately the Dimotakis speed $U_{D} = \sqrt{r}/\left( 1 + \sqrt{r} \right)$, which is a relevant measure of the propagation speed of fully-developed disturbances in $2D$ mixing layers \citep{Dimotakis_1986}. By following the center of recirculation vortices downstream of the wave, their trajectory is drawn as solid curved lines with large dots (blue online) in figure \ref{fig:BagBreakUpSimu}. We note that the vortices are shed periodically and each shedding event coincides with a droplet ejection event. Figure \ref{fig:BagBreakUpSimu} displays three such droplet ejection events at $t = 50-90$, $t = 90-140$ and $t = 150-200$.
	
	We now point to figure \ref{fig:ExpeVortexShedding}c where one complete droplet ejection event from these simulations (corresponding to $r=0.02$) is displayed side by side with that of the experimental images. Here, the time evolution of the gas--liquid interface and flow streamlines is shown as thick mid-gray lines (red online) and thin gray lines, respectively. Even though the simulation pertains to a rather different flow configuration at a moderate density ratio (a single nonlinear wave in a large periodic domain with a density ratio twenty times larger), we can nevertheless recognise the same \textit{droplet catapult sequence}, leading once again to violent ejection of liquid droplets. The sequence is very similar to that of the air-water mixing layer experiments in figures \ref{fig:ExpeVortexShedding} a--b in \S\ref{sec:VortexSheddingExpeObserv}.  Supplementary videos \cite{SuppMovie} are provided where $4$ such vortex shedding events and subsequent droplet catapult processes are seen in the numerical simulations (see \textit{Movie 5} \cite{SuppMovie}).
\begin{figure}[t]
\begin{center}

\epsfig{file=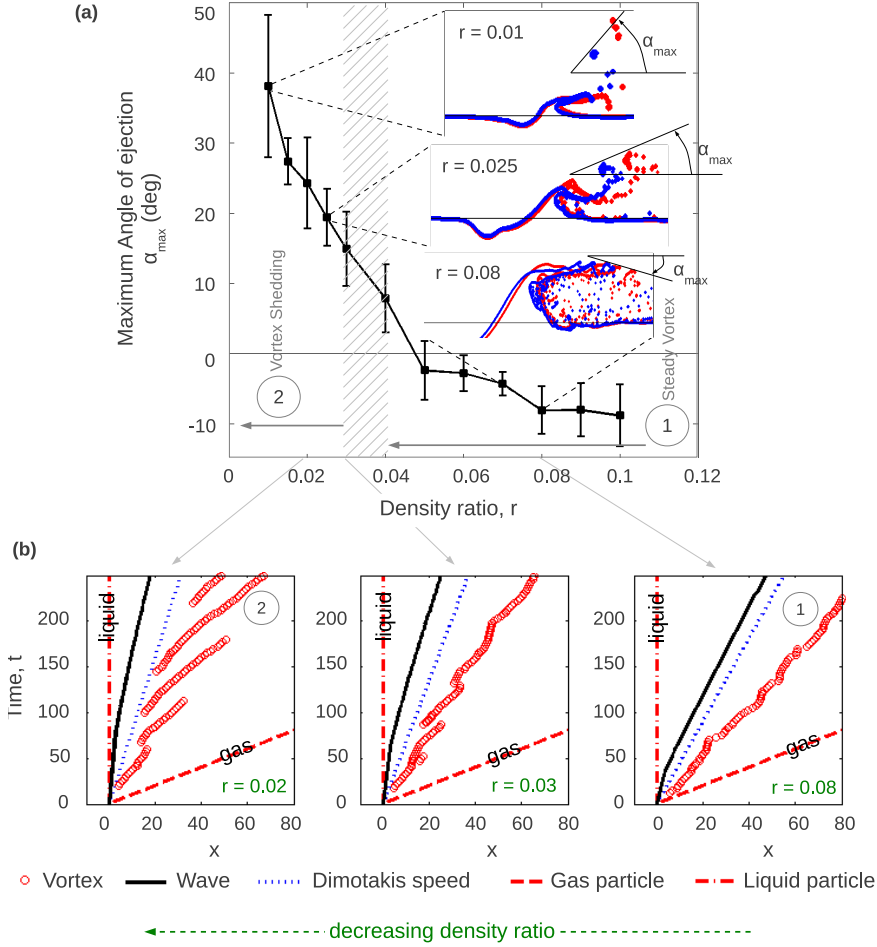,width=0.7\textwidth,keepaspectratio=true,bbllx=145,bblly=391,bburx=445,bbury=717,clip=}

\end{center}
\caption{(a) Maximum droplet ejection angle $\alpha_{max}$ obtained from $DNS$ as a function of density ratio $r = \rho_{g}/\rho_{l}$. It displays a sharp increase to positive values at $r \leq 0.04$. (b) Streamwise evolution of the wave and vortex centres with time. Multiple vortex shedding events are observed when $r < 0.04$.}
\label{fig:EjectionAngle}
\end{figure}

\begin{figure}
\begin{center}

\epsfig{file=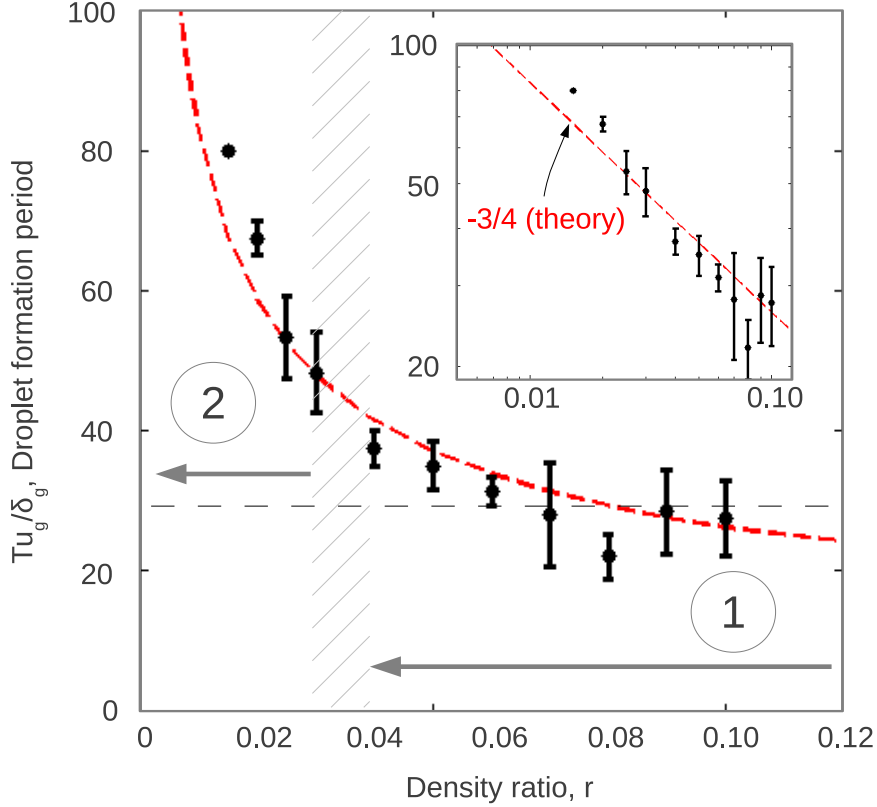,width=0.7\textwidth,keepaspectratio=true,bbllx=145,bblly=433,bburx=450,bbury=716,clip=}

\end{center}
\caption{Variation of droplet formation period $TU_{g}/\delta_{g}$ with density ratio $r = \rho_{g}/\rho_{l}$ as obtained from $DNS$. A sudden increase in the time taken to form droplets is observed as the droplet catapult mechanism sets-in. The inset displays the same graph on a log-log scale comparing the theoretical exponent ({\color{red}$---$}) with the data.}
\label{fig:EjectionPeriod}
\end{figure}

Figures \ref{fig:ExpeVortexShedding}a--c already indicate that vortex shedding is connected to droplet ejection processes. A quantitative measure of the effect of this change in gas flow dynamics on droplet ejection angles can be deduced by varying the density ratio between the gas and the liquid. From \citet{Hoepffner_2011}, we know that the density ratio affects very much the morphology of the wave: at $r=1$ the self-similar wave is symmetric with respect to its center and it is composed of two large vortices; decreasing $r$ progressively, the wave loses its symmetry and takes the shape of a liquid body upstream of a gas vortex. Thus, by following the processes of droplet creation while decreasing $r$ progressively, we may shed light upon the connection between vortex shedding and the catapult mechanism.

Figure \ref{fig:EjectionAngle}a displays the variation of maximum droplet ejection angle $\alpha_{max}$ over density ratio $r$. The error bars quantify the standard error over successive ejection events for a given density ratio. This angle $\alpha_{max}$ is computed by superposing snapshots of the interface for two consecutive time units. It is then given by the maximum angle that the superposed droplets make with the streamwise direction as shown, for example, in the insets of figure \ref{fig:EjectionAngle}a corresponding to $r = 0.08$, $0.025$ and $0.01$. As the density ratio $r$ decreases, $\alpha_{max}$ initially remains approximately constant and negative until $r \approx 0.04$; a negative ejection angle implies that the drops fall downward towards the interface. Thus, there is no droplet catapult when $r > 0.04$. When the density ratio is decreased further, there is a steep increase in the angle of ejection, and $\alpha_{max}$ as high as $\approx 40^\circ$ is observed. A linear extrapolation of the angle of ejection data predicts $\approx 50^\circ$ for the case of air-water. This prediction matches the maximum angle of ejection (see figure \ref{fig:AngleExperimental}a) obtained in the experiments of \citet{Raynal_thesis_1997}.

	Figure \ref{fig:EjectionAngle}b presents the spatio-temporal evolution of the wave/vortex system for various density ratios. Here, the wave centre (thick continuous line) refers to the location where the nonlinear $KH$ wave crosses the centreline $y = 0$. The wave and vortex centres are determined via manual inspection of the interface and the flow streamlines from the beginning to the end of the simulation. As expected from \citet{Hoepffner_2011}, the wave center moves downstream at a constant speed close to the Dimotakis speed $U_D$ (dotted line). As for the vortex motion (denoted by {\color{red}$\circ$}), we may distinguish two regimes: regime \circled{1} for $r>0.04$ where the speed of the vortex is constant, it remains attached to the wake of the wave, and regime \circled{2} for $r<0.03$ where the wave is very slow and its recirculation region is unstable: the vortex is shed and is periodically replaced by new vortices.
	
	In both regimes \circled{1} and \circled{2}, the thin liquid film oscillates and breaks-up into droplets. During one complete flapping cycle, the liquid film displaces first vertically downward and then upward. The droplets are formed at the end of each flapping cycle. This implies that the time between ejection events is equal to the flapping period. In regime \circled{2}, numerical simulations show that the flapping motion and eventual break-up of the liquid film is synchronised with the vortex shedding process (see for example, \textit{Movie 5} \cite{SuppMovie}). Therefore, in this case, the droplet ejection period, the flapping period and the vortex shedding period are the same.
	
	Figure \ref{fig:EjectionPeriod} displays these characteristic time periods as a function of density ratio $r$. It is measured by observing the time evolution of the  gas-liquid interface and droplets in the self-similar coordinates, namely, $x^{'} = x/U_{g}t$ and $y^{'} = y/U_{g}t$. The measured period is taken to be the time between newly formed droplets to cross the line $x^{'} = x/U_{g}t = constant$. The errorbars quantify the standard error between each such events at a given density ratio. As the density decreases from $r = 0.12$, the droplet formation period remains more or less constant. However, at $r \leq 0.03$, it increases rapidly. The rapid increase in droplet ejection period coincides with the onset of vortex shedding.
	
	In a backward-facing step, the vortex shedding period is independent of the fluid density. In the case of a localized $KH$ wave, however, the wave height (which represents the characteristic length scale of the obstacle) depends on the density ratio \citep{Hoepffner_2011}. Hence, it is not surprising that the measured period $TU_{g}/\delta_{g}$ depends on the density ratio.
	 
	If  $a_{d}$ denotes the acceleration due to the aerodynamic force $F_{d}$ experienced by a thin liquid film of mass $m_{f}$, we have
\begin{align}
a_{d} &= \frac{F_{d}}{m_{f}}, \\ \notag
a_{d} &= \frac{\frac{1}{2} C_{d}\left( \rho_{g} U_{g}^2 A_{f} \right)}{\rho_{l}(A_{f} \times b)},
\label{eq:acceleration}
\end{align}
where $C_{d}$ is the coefficient of drag, $b$ is the thickness of the film, $A_{f}$ is the projected frontal area of the film and it is equal to the length of the film times its width. The thickness of the film $b$ can be taken as proportional to the fastest growing wavelength of the Kelvin-Helmholtz instability $\sim \delta_{g}/\sqrt{r}$ \citep{Rayleigh_1879,Drazin_n_Reid_1981,Hong_2003, Marmottant_2004}. The aerodynamic acceleration of the film is $\mathcal{O}(\Delta l/T^{2})$, where $\Delta l$ is the distance covered by the liquid film in time $T$ before it breaks into droplets. Since a liquid film breaks when the aerodynamic pressure due to the recirculation region is too large to be supported by the surface tension forces on the liquid film, $\Delta l$ should depend only on the Weber number. So, as a first approximation, it is a constant with respect to the density ratio $r$ and hence, we obtain
\begin{equation}
\frac{T U_{g}}{\delta_{g}} \propto r^{-3/4},
\label{eq:droplet_formation_time}
\end{equation}
where the proportionality constant depends on the gas Reynolds number via $C_{d}$ and the gas Weber number. The inset of the figure \ref{fig:EjectionPeriod} compares this prediction with $DNS$ computations. We observe that the theoretical exponent $-3/4$ based on aerodynamic force argument shows an overall agreement with the time taken for droplet formation in the simulations.

\section{Discussion}
\label{sec:Discuss}
	We have, thus, identified the \textit{droplet catapult mechanism} in two configurations, namely, the splitter plate experiment and numerical simulations. In the following, we first, briefly point out how the simulations differ from the experimental set-up. Using videos from high-speed camera imaging, we comment on the variation of $\alpha_{max}$ with air and water speed in experiments. Thereby, we provide an explanation for the results of \citet{Raynal_thesis_1997} (figure \ref{fig:AngleExperimental}a).
	
\subsection{Difference between our splitter plate experiments and $DNS$ computations}
\label{subsec:Discuss1}
	
	We examined the \textit{droplet catapult mechanism} in a relatively simple flow situation consisting of a nonlinear localized Kelvin-Helmholtz wave in two-phase mixing layers. Direct Numerical Simulations allowed us to readily extract quantitative information. The experimental flow is a spatially-developing air-water shear layer wherein large liquid waves are periodically formed at the trailing edge of a splitter plate, whereas the simulations correspond to the spatio-temporal evolution of an infinite $2D$ shear layer excited initially by a localised impulse. Here, the only control parameter is the density ratio $r$.
 
	In addition, a single nonlinear wave performs the catapult sequence repeatedly (see figure \ref{fig:BagBreakUpSimu}): it is not simply one event per wave as in the case of splitter plate experiments. There are, in general, as many successive events per wave as the computational box can afford (see for example, \textit{supplementary video 5}  \cite{SuppMovie}). The process ends only when the wave has grown to an extent when the computational box becomes too small compared to its size. Note that a few instances of successive catapult sequences on a single wave can also be observed in the videos (\textit{Movie 3} \cite{SuppMovie}) from experiments when the air speed is sufficiently large. In the experiments, waves are formed periodically at the trailing edge of the splitter plate and so, the first wave is soon shadowed by the appearance of a nascent wave at the trailing edge. Thus, the first wave loses its wind and thus becomes a collapsing passive wave.

\subsection{Effect of air and water speed on $\alpha_{max}$.}
\label{subsec:Discuss2}

\begin{figure}[t]
\begin{center}

\epsfig{file=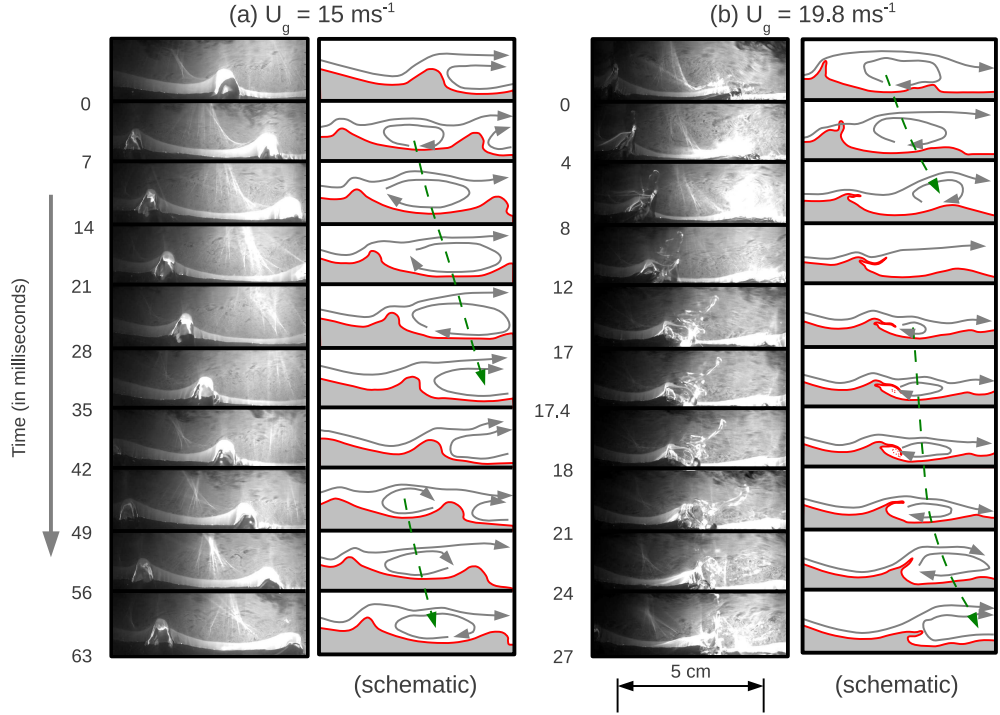,width=0.8\textwidth,keepaspectratio=true,bbllx=144,bblly=489,bburx=462,bbury=718,clip=}

\end{center}
\caption{The sequence of interfacial patterns for various air speeds $U_{g}$ at a constant water speed $U_{l} = 0.23 ms^{-1}$ from planar two-phase mixing layer experiments: (a) $U_{g} = 15$ $ms^{-1}$ and (b) $U_{g} = 19.8$ $ms^{-1}$. For the sake of clarity, schematics of the snapshots are also presented in both figures. For a wider range of air speeds see \textit{supplementary videos 3} \& \textit{4}  \cite{SuppMovie}. At these air speeds vortex shedding does not occur and droplet catapult mechanism is absent.}
\label{fig:ExpeUgasSequence}
\end{figure}

	In the case of experiments, our observations (figure \ref{fig:ExpeVortexShedding}a--b) in section \S\ref{sec:VortexSheddingExpeObserv} correspond to the air speed $U_{g} = 25.5  ms^{-1}$. This is approximately the speed when the maximum angle of ejection $\alpha_{max}$ is the largest (see figure \ref{fig:AngleExperimental}). At the same water speed $U_{l} = 0.23 ms^{-1}$, if the air speed is smaller we find that the droplet catapult mechanism is absent. For instance, consider the case displayed in figure \ref{fig:ExpeUgasSequence}a where $U_{g} = 15  ms^{-1}$. We observe waves in the form of small bumps that appear periodically at the splitter plate and move progressively downstream at a constant speed. At this air speed, the liquid film does not form from the crest of the wave. The air flow over the air-water interface shows the presence of a recirculation region. However, the recirculation region remains attached to the wave and moves at approximately same speed as the wave. 

	Similarly, the case corresponding to a slightly larger air speed ($19.8  ms^{-1}$) is shown in figure \ref{fig:ExpeUgasSequence}b. In contrast with the case in figure \ref{fig:ExpeUgasSequence}a, the air-water interface wave forms a thin liquid film. The incoming air flow and the recirculation zone act on the liquid film. This results in the up and down motion of the film that is observed in figure \ref{fig:ExpeUgasSequence}b. Nevertheless, after the first vortex shedding event, the second recirculation vortex remains steady. We observe that this vortex does not cause \textit{bag-breakup from below}. Instead, the liquid film forms a hole that develops and ruptures as seen for $17-18 \times 10^{-3} s$ in figure \ref{fig:ExpeUgasSequence}b.

	At the same water speed $U_{l} = 0.23  ms^{-1}$, if the air speed is much faster ($U_{g} > 25.5  ms^{-1}$) compared to that in figures \ref{fig:ExpeVortexShedding} \& \ref{fig:ExpeUgasSequence}, the wave originating from the splitter plate is much smaller and forms droplets close to the trailing edge of the plate. Since still images from the experiments are not very easy to interpret, we provide the viewer with \textit{Movies 3 $\&$ 4} \cite{SuppMovie}. They correspond to airspeeds of $30 ms^{-1}$ and $32.5 ms^{-1}$. Instead of the droplet catapult sequence, we observe that the liquid film suddenly bursts to form droplets. There is not enough time for the droplet catapult sequence to be performed step by step to project the resulting liquid drops at a large angle. Since this happens when the wave is small, the film is very thin. So, the quantity of water that can be ejected is small as well. The reader is referred to \textit{supplementary videos 3} \& \textit{4} where the air flow visualization is clearer (a wider range of air velocities is presented in these videos) \cite{SuppMovie}.
	
	These observations imply that, at a given liquid speed, if the wind is low, a steady recirculation zone is formed in the wake of the wave. In this case, the liquid film is pushed by the incoming air flow and is trapped in the recirculation region. It eventually segments into droplets that are not catapulted by the recirculation vortex. This corresponds to a negative ejection angle. The situation resembles the case of our simulations for density ratio $r=0.08$ (and above) as shown in the inset of figure \ref{fig:EjectionAngle}a. Now, if we progressively increase the wind speed, the recirculation region becomes unstable and vortex shedding occurs. Figure \ref{fig:AngleExperimental}a shows that this starts to happen at about $U_{g} = 20 ms^{-1}$. This is the value of the wind speed at which we start to be able to measure a drop ejection angle. From this value the maximum ejection angle increases quickly, until a maximum for a gas speed of about $25ms^{-1}$. At larger air speeds, however, the ejection angle is smaller and it decreases progressively with $U_{g}$. This explains the data from Raynal's experiments\cite{Raynal_thesis_1997}, as presented above in figure \ref{fig:AngleExperimental}a, for water speed up to $1.11ms^{-1}$.

	However, for $1.11ms^{-1}$ the first peak is lost and we see a much flatter peak at an air velocity of about $50ms^{-1}$. We have not studied this regime in detail, but we observed that when the liquid velocity is large, the shape of the waves and their frequency change. The departure of each nonlinear wave carries a large mass of water away from the trailing edge of the splitter plate. This leads to a local depletion, and a new wave can be born only once the stream is refilled from the water inlet supply. Thus, for slow liquid, the waves are fewer and larger, which leaves plenty of time for the catapult sequence to play its role, whereas for fast liquid, the waves are many and smaller, with hardly any time for the synchronised sequence.


\section {Conclusion}
	\label{sec:Conclu}

We studied the process of \emph{droplet catapult} in two-phase mixing layer via experiments and simulations. First, a nonlinear wave grows under the Kelvin--Helmholtz instability. This wave moves slowly downstream and is an obstacle to the gas stream. Thus, a recirculation region appears in wake of the wave. Because of the gas shear on top of the wave, a thin liquid film emerges from the crest of the wave. The fate of this liquid film now depends on the behaviour of the recirculation zone: if the recirculation remains attached to the wave (fast wave or slow gas), the film is trapped into the stationary recirculation zone and breaks-up into drops that fall towards the gas-liquid interface. This is the case in the splitter plate experiment for low air speeds and, also, in the numerical simulation for density ratios  $r$ close to one. If on the other hand, the recirculation is unstable and vortex shedding occurs, we observe the catapult sequence due to a synchronised motion of the liquid film and the gas flow streamlines. This sequence is itemized in the abstract of this article. The liquid film is very thin and hence, it is advected by the evolution of gas flow configuration of the departing vortex, as observed on figure \ref{fig:ExpeVortexShedding}c.  The departure of the vortex implies a momentary reattachment of the gas flow and eventually, the formation of a new recirculation vortex. This new vortex blows-up the liquid film such that the droplet ejection occurs via the violent event of \emph{bag-breakup from below}.

	Ejection angles in mixing layers provide only a hint of where the liquid stripped from the perturbed mixing layer is sent. On the other hand, droplet size distribution is a more precise and useful quantity in order to completely quantify the dispersed two-phase flow. It is expected that our present study would provide more insight on further research in that direction. \citet{Orazzo_2012} showed that the evolution of localized Kelvin-Helmholtz wave in the presence of gravity is relevant to spontaneous creation of large oceanic waves. In this context, our results imply that air recirculation zones could influence spray formation. But, in general, large oceanic waves are far from fully developed Kelvin-Helmholtz waves and hence, our conclusions cannot be directly translated to such situations.

J. J. S. J. and J. Hoepffner acknowledge Gilles Agbaglah, Daniel Fuster, Pierre-Yves Lagr\'{e}e and Pascal Ray of the Institut D'Alembert for useful discussions on using Gerris. We also thank Antoine Delon's kind assistance during our visit to LEGI. 

This project has been supported by the ANR VAA (ANR-2010-BLAN-0903) program, the ANR DYNAA (ANR-2005-BLAN-0213) program and the FIRST (Fuel Injector Research for Sustainable Transport) project supported by the European Commission under the $7$th Framework Programme.

\end{document}